\documentclass{article}  
\usepackage{latexsym,amssymb,amsthm}
\usepackage[dvips]{graphicx} 
\usepackage{multirow}
\usepackage{color}

\AtBeginDvi{}

\title{\LARGE \bf 
Match results prediction ability of official ATP singles ranking
}

\author{Eiji Konaka$^{a}$
\thanks{$^{a}$Meijo University.
1-501, Shiogamaguchi, Tenpaku-ku, Nagoya, JAPAN.
       {\tt\small konaka@meijo-u.ac.jp}}%
}

\begin{document}
\maketitle

\begin{abstract}
This paper discusses match result prediction ability of ATP ranking points, which is official ranking points for men's professional tennis players.
The structure of overall ATP World Tour and the ranking point attribution system leads that the ranking point ratio between two players is an essential variable.
The match result prediction model is a logistic model.
The fact is verified using statistics of over 24000 matches from 2009.
\end{abstract}

\section{Introduction}

Match result prediction model has much attention in recent years.
The professional tennis is no exception.
The paper \cite{jqas-2015-0059} reviews several prediction models (regression model, pointwise prediction, paired-comparison, and so on), and reports match result prediction ability for each methods.

Original prediction models have been extensively reported. For instance; 
\begin{itemize}
\item Utilizing ranking difference 
\cite{9c3cc2c35fb7409ab50def4c7115a0aa}.
\item Google PageRank\cite{Brin:1998:ALH:297810.297827} is applied to tennis 

\cite{Dingle2013}. 
\item Calculate scoring probability during the match \cite{Klaassen2003257}.
\item Pairwise comparison problem is formulated as matrix-based method 
\cite{DAHL201226}.
\item Bradly-Terry model \cite{10.2307/2334029} with effect of court surfaces
\cite{McHale2011619}.
\end{itemize}

Surprisingly, there is no report on the statistical properties and the match prediction ability of the official ATP (Association of Tennis Professionals) ranking points.
If it is mentioned, the very simple prediction ``the player with higher ranking will win" is compared to the one proposed by the authors of each paper.
Some authors uses the ranking gap and/or ranking point gap as explanatory variables in regression model to make their prediction model.
In these cases, the qualitative meaning oｆ the obtained coefficient and the model itself have not been fully explained.

The objective of this paper is to analyze how the official ranking point can explain the match result.

The main result of this paper is as follows:

\begin{itemize}
\item The ATP ranking, the official ranking point for men's professional tennis, can predict match result probability via logistic function with a ranking point ratio between two players as a variable. 
\item The match prediction ability is based on consistent tour structure design and point attribution system.
\end{itemize}


\section{Design of ATP ranking points }
This section reviews the design of ATP ranking points.
A hypothesis that the ratio of the ranking points is essential can be derived from their definitions.

\subsection{ATP world tour and  ATP rankings}

ATP renovates overall tour structure in 2009.
The ranking point system is also re-designed at the same time.
Professional tennis players (member of ATP) participate in several tournaments hosted in various places all over the world.
The series of tournaments are called as ``ATP world tour".

Each tournament is classified into one of the tournament categories.
Each tournament category has different size of draws, prize money, and ATP ranking points. Large and famous tournaments, e.g., four Grand Slams, have large ranking points.
Table \ref{tab:tournamentCategories} lists the main tournament categories of ATP world tour in 2017.
\begin{table}[h]
\caption{Tournament categories}
\label{tab:tournamentCategories}
\begin{center}
\begin{tabular}{p{2cm}p{2cm}p{2cm}p{2cm}p{1.5cm}}\hline
Category name& Number of tournaments & ATP ranking points for winner&
Draws (tournaments)&Governing body\\ \hline \hline
Gland Slam&4&2000&128(4)&ITF \\ \hline
ATP World Tour Masters 1000&9&1000&96(2), 56(6), 48(1)&ATP \\ \hline
ATP World Tour  500&13&500&48(2), 32(11)&ATP \\ \hline
ATP World Tour  250&40&250&48(1), 32(2), 28(37)&ATP \\ \hline \hline
ATP Finals&1&1100 -- 1500&8 (1)&ATP \\ \hline
\end{tabular}
\end{center}
\end{table}

ATP ranking points are awarded also in tournaments such as ATP challenger tour and ITF\footnote{International Tennis Federation.} futures series, not listed in Table \ref{tab:tournamentCategories}. 

A player's ATP ranking points is basically based on the total points he accrued in the following 18 tournaments in past 52 weeks.
Top 30 players in the last year have to participate in four Grand Slams and 8 Masters tournaments. 
The detail of exceptions are not mentioned in this paper. Please refer \cite{http://www.atpworldtour.com/en/rankings/rankings-faq}.



\subsection{Design of ATP ranking points and tournament structure}

Table \ref{tab:tournamentCategories} shows that the ratio of winner's ranking points between adjacent tournament categories are equal, i.e., the winner's ranking points are 250, 500, 1000, and 2000, doubled with climbing up one category.  
Table \ref{tab:tournamentCategoriesRankingPoints} shows the detailed ranking point for each result. One win in each tournament shall increase ranking points in double or 5/3 times.

\begin{table}[h]
\caption{Ranking points in each tournament categories \cite{http://www.atpworldtour.com/en/rankings/rankings-faq}}
\label{tab:tournamentCategoriesRankingPoints}
\begin{center}
\begin{tabular}{p{2.2cm}ccccccccc}\hline
Category name& W&F&SF&QF&R16&R32&R64&R128&Q\\ \hline \hline
Gland Slam&2000&1200&720&360&180&90&45&10&25\\ \hline
ATP World Tour Masters 1000&1000&600&360&180&90&45&10(25)&(10)&16\\ \hline
ATP World Tour  500&500&300&180&90&45&(20)&&&20\\ \hline
ATP World Tour  250&250&150&90&45&20&(5)&&&12\\ \hline
\end{tabular}
\end{center}
\end{table}

Top players are seeded so that these players do not match in earlier stages.
Explain 32 draws tournament with 8 seeded players\cite{atpRulebook-PlacementOfSeeds}．

\begin{itemize}
\item Make simple tournament bracket by 32 players. Each place are numbered as $1, 2, \cdots, 32$ from the top.

\item 8 seeded players are determined based on the ATP ranking.
1st and 2nd seeded players are placed in numbers 1 and 32. 
\item 3rd to 8th seeded players are grouped as follows:
\begin{itemize}
\item 3rd and 4th seed: 9, or 24.
\item 5th to 8th see:  8, 16, 17, or 25.
\end{itemize}
Placements in each group are determined by ballots.
\end{itemize}
As a result, 1st and 2nd, 1st to 4th, and 1st to 8th seeded players are not matched before Final (F), Semifinals (SF), and Quarter-finals (QF).
If the draws are larger than 32, e.g., 64 or 128, the placements of the seeded players are determined in the similar manner.

\subsection{Hypothesis on relation between raking point and winning probability}
The following hypothesis can be stated based on before mentioned consistent design of the ranking point, mandatory tournaments for top players, and seed players' placement design for each tournament.

{\bf Hypothesis：}

\begin{itemize}
\item
The logarithm of the ranking point shows how high the player can climb up in tournament bracket because the attributed ranking point grows almost the same rate with one victory. 
\item At the same time, the gap of the ability of two players is a function of the ratio of their ranking points.
	\begin{itemize}
	\item 
The predicted winning probability converges to 1 when the ratio is large.
The value is expected to be 0.5 if the ratio is 1.
Therefore, the predicted winning probability can be modeled by the logistic function 
\begin{equation}
\displaystyle p(x)=\frac{x^\alpha}{1+x^\alpha},~~ \alpha>0.
\end{equation}
	\end{itemize}
\item Top 30 players in the last year have to participate in four Grand Slams and 8 Masters tournaments. 
They have to choice at least 6 tournaments from 500 or 250 series because the ranking point is the sum of the points earned in 18 tournaments. 	\begin{itemize}
	\item The grater part of the ranking of the top players are determined by the result of the Grans Slams and Masters tournaments because all top players plays in these tournaments. 	
	\item It is not rational to participate 6 (or more) 500 series tournaments because the other top players also will participate in these tournaments. 
	\item 
If every top 30 players participate in 3 500 series tournaments, each 500 series tournament has 7 players out of top 30 players in average because 13 500 series tournaments are held in one year.
Therefore, QF in 500 series tournament are equivalent to R32 in Grand Slams.
Their ranking points are equal to 90. 
	\item If every top 30 players participate in 3 250 series tournaments, each 250 series tournament has 2.5 players out of top 30 players in average because 40 250 series tournaments are held in one year.
Therefore, SF in 250 series tournament are equivalent to R32 in Grand Slams.
Their ranking points are equal to 90.
 	\end{itemize}

Based on these insights, (absolute) ranking and expected ranking points in each tournament category are listed in Table \ref{tab:rankingAndExpectedRankingPoints}.


\begin{table}[h]
\caption{Ranking and expected ranking points}
\label{tab:rankingAndExpectedRankingPoints}
\begin{center}
\begin{tabular}{lccccccccc}\hline
Ranking&  & & & &16&32&64\\ \hline \hline
Gland Slam (4)&2000&1200&720&360&180&90&45\\ \hline
ATP World Tour Masters 1000 (8)&1000&600&360&180&90&45&25\\ \hline
ATP World Tour  500 (3)&&&500&300&180&90&45\\ \hline
ATP World Tour  250 (3)&&&&250&150&90&45\\ \hline \hline
Expected ranking points& &&&&2430&1260&650 \\ \hline
\end{tabular}
\end{center}
\end{table}

An ideal player at ranking 32 will gain $90\times 4 + 45\times 8 + 90 \times 3 + 90 \times 3 =1260$ ranking points.
The other ideal players at rankings 16 and 64 will gain $2430$ and $650$, based on the similar discussions.
The $2430$ shall give an estimate of the upper bound for the ideal player because the player at rank 16 cannot always win against the player at rank 32.
On the other hand, the $650$ shall give an estimate of the lower bound for the ideal player because lower ranked players can obtain ranking points in the lower categories such as ATP challenger tours.

\end{itemize}

This hypothesis will be verified below.
\section{Analysis result}

This section shows analysis result based on the ranking data disclose by ATP, and the database maintained by \cite{https://github.com/JeffSackmann}.

\subsection{Number of participating tournaments and ranking points}

\begin{figure}[h]
\begin{center}
	\includegraphics[width=1.0\columnwidth ,clip]{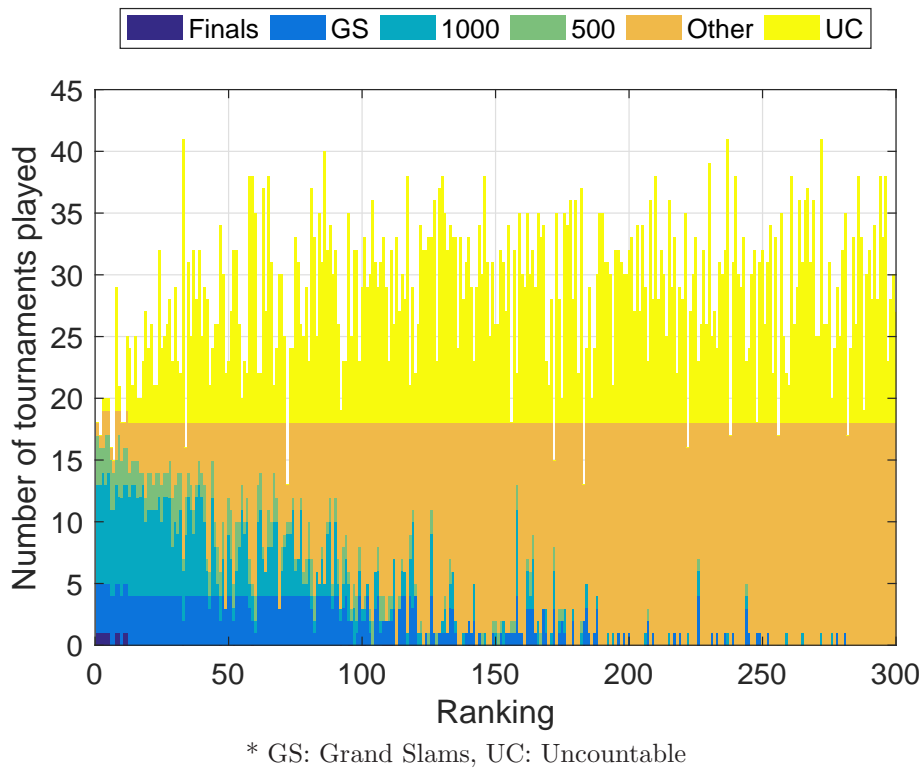}
* GS: Grand Slams, UC: Uncountable
	\caption{Number of tournaments played in previous 52 weeks}
	\label{fig:tournamentNumbers}
\end{center}
\end{figure}

\begin{figure}[h]
\begin{center}
	\includegraphics[width=1.0\columnwidth ,clip]{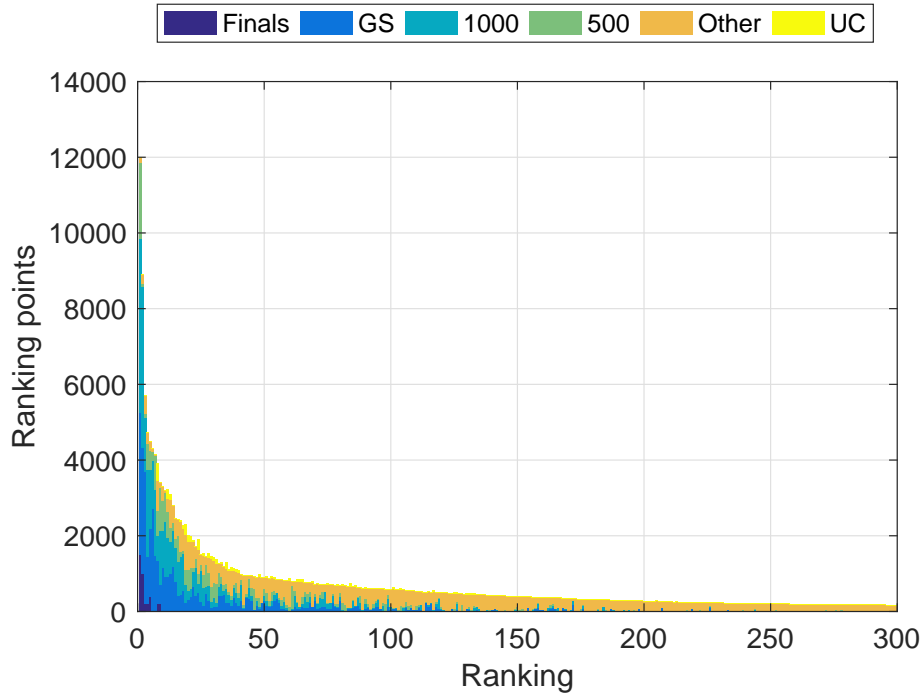}
* GS: Grand Slams, UC: Uncountable
	\caption{Ranking points}
	\label{fig:points}
\end{center}
\end{figure}

\begin{figure}[h]
\begin{center}
	\includegraphics[width=1.0\columnwidth ,clip]{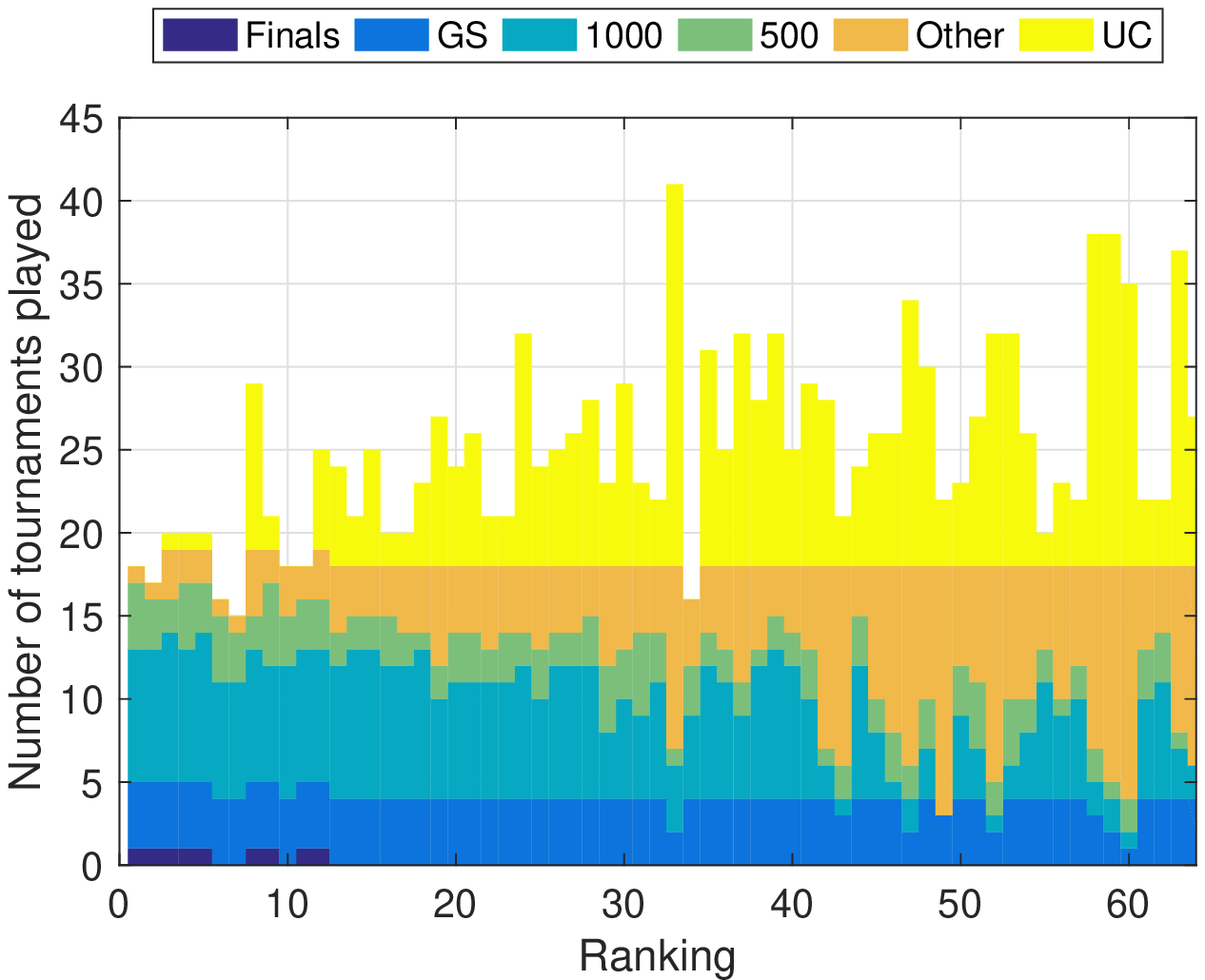}
* GS: Grand Slams, UC: Uncountable
	\caption{Number of tournaments played in previous 52 weeks (top 64 players)}
	\label{fig:tournamentNumbersTop}
\end{center}
\end{figure}

\begin{figure}[h]
\begin{center}
	\includegraphics[width=1.0\columnwidth ,clip]{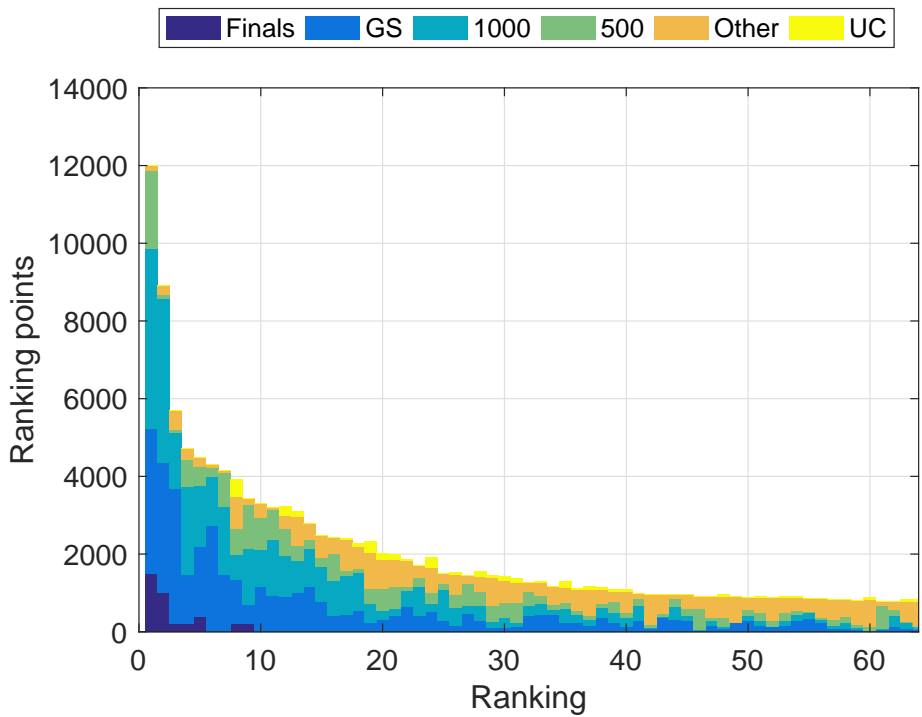}
* GS: Grand Slams, UC: Uncountable
	\caption{Ranking points (top 64 players)}
	\label{fig:pointsTop}
\end{center}
\end{figure}

Top 300 players in the official ATP ranking \cite{atpRanking20170320} are collected.
Figures \ref{fig:tournamentNumbers} and \ref{fig:points} show the number of tournaments and the ranking points divided into the tournament categories.
Figures \ref{fig:tournamentNumbersTop} and \ref{fig:pointsTop} show top 64 players.

Table \ref{tab:500250} shows the number of participating tournaments in 500 or 250 series for top ranked players.

\begin{table}[h]
\caption{Number of ATP 500 and 250 tournaments played by top players}
\label{tab:500250}
\begin{center}
\begin{tabular}{cccccccccc}\hline
&&0&1&2&3&4&5&6 or more& Mean\\ \hline \hline
\multirow{2}{30pt}{Top 8}&ATP World Tour 500
	& 0 &	0  &   2   &  3  &   3  &   0  &   0  &   $3.125$\\ 
&ATP World Tour 250
	& 0 &  4  &   2   &  1   &  1  &   0   &  0  &   $1.875$\\ \hline
\multirow{2}{30pt}{Top 16}&ATP World Tour 500
	& 0 &	0 &    5    & 7  &   3  &   1   &  0& $3.000$\\ 
&ATP World Tour 250
	& 0 &  4   &  4   &  6   &  2   &  0   &  0&$2.375$\\ \hline
\multirow{2}{30pt}{Top 30}&ATP World Tour 500
	& 0 & 1   & 11 &   13  &   4    & 1  &   0  & $2.767$ \\ 
&ATP World Tour 250
	& 0 &  4   &  4   &  7   & 10    & 3  &   2  & $3.333$\\ \hline
\multirow{2}{30pt}{Top 64}&ATP World Tour 500
	& 2   &  7   & 25   & 22   &  6   &  2    & 0  &$2.453$\\ 
&ATP World Tour 250
	& 0   &  4    & 4   &  9  &  16  &   8 &   23 & $5.703$\\ \hline
\end{tabular}
\end{center}
\end{table}
As expected in the previous section, the top 30 players participates 3 tournaments  for both 500 and 250 series in average.
The higher ranked players participate more 500 tournaments, on the other hand, lower ranked players participate more 250 series.

\begin{figure}[h]
\begin{center}
	\includegraphics[width=1.0\columnwidth ,clip]{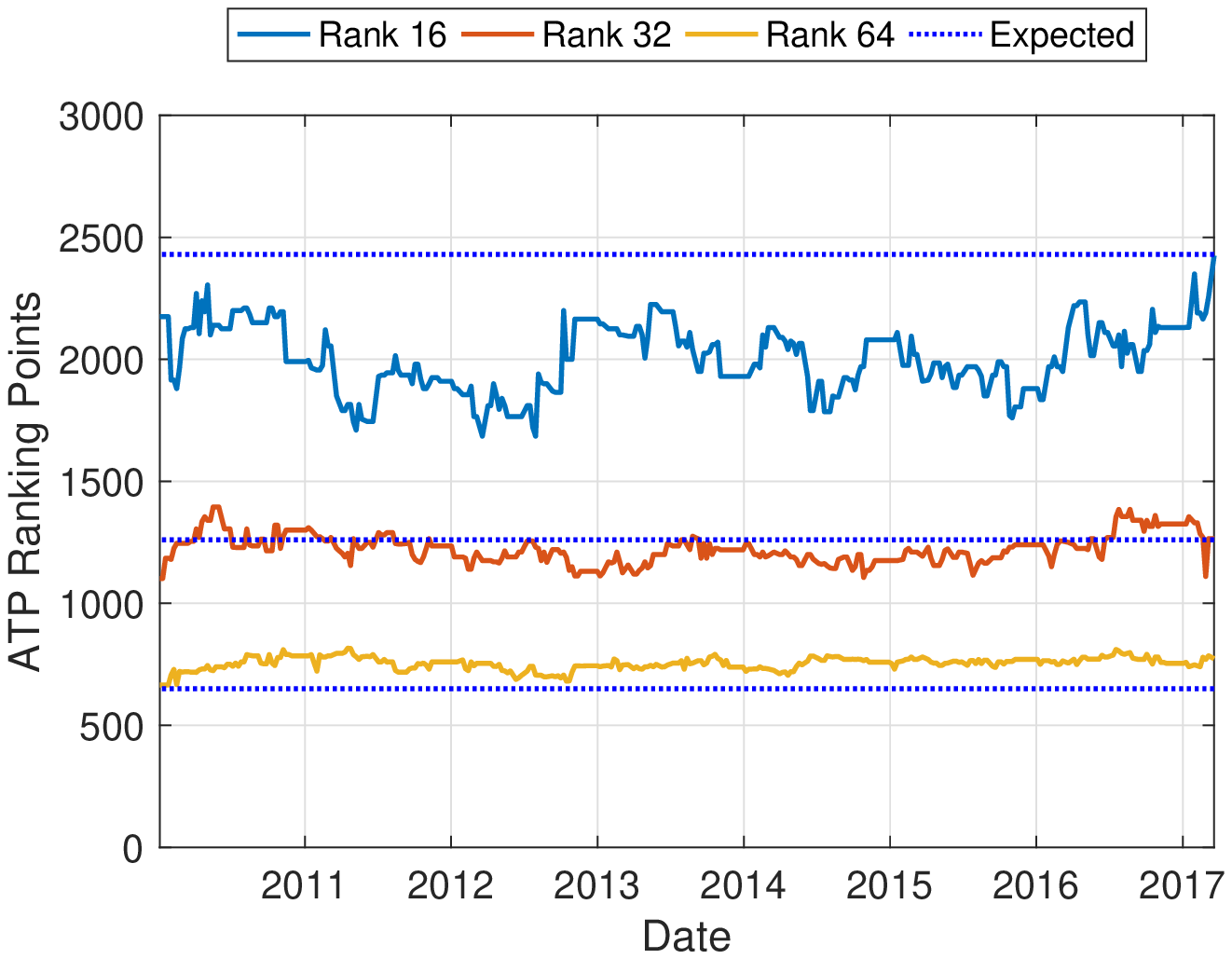}
\caption{Ranking points at rank 16, 32, and 64}
	\label{fig:rankingPoints16-32-64}
\end{center}
\end{figure}

Figure \ref{fig:rankingPoints16-32-64} shows the ranking points at rank 16, 32, and 64 from 2010.
Table \ref{tab:rankingPoints16-32-64} lists some basic statistics.

\begin{table}[h]
\caption{Ranking points at rank 16, 32, and 64}
\label{tab:rankingPoints16-32-64}
\begin{center}
\begin{tabular}{cccc}\hline
& $16$ &$32$ &$64$ \\ \hline \hline
Expected&{\bf 2430}&{\bf 1260}&{\bf 650}\\ \hline
2017.03.20&2425&1265&773 \\ \hline
Maximum&{\bf 2425}       & 1395        & 816\\ \hline
Mean& 2009.8   & {\bf 1224.4}    &753.5\\ \hline
Minimum&1685        &1102        & {\bf 665}\\ \hline
Std$*$&138.40   &61.04   &26.38\\ \hline
\end{tabular}
\vspace{5pt}

$*$ Std: standard deviation
\end{center}
\end{table}

This result shows that the hypothesis,
\begin{itemize}
\item ranking point $2430$ gives an upper bound for rank 16,
\item ranking point $1260$ gives an estimate for rank 32, and
\item ranking point $650$ gives an lower bound for rank 64,
\end{itemize}
can not be rejected, and it has no huge mistakes.
In particular, 1260 points for rank 32 is an useful estimate.

\begin{figure}[h]
\begin{center}
	\includegraphics[width=1.0\columnwidth ,clip]{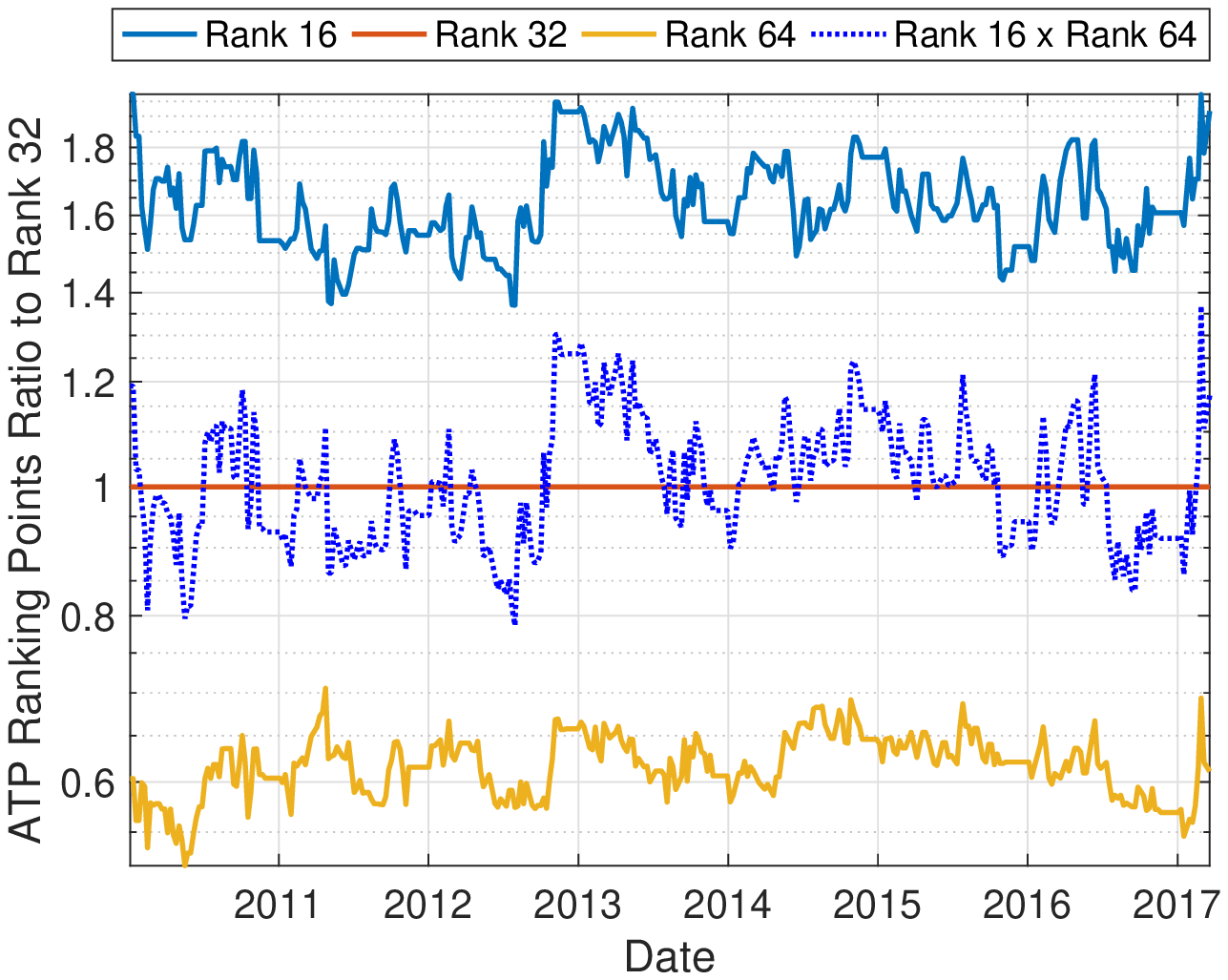}
\caption{Ranking points ratio to rank 32}
	\label{fig:rankingPoints16-32-64RatioTo32}
\end{center}
\end{figure}

Figure \ref{fig:rankingPoints16-32-64RatioTo32} shows the ratio to the ranking point at rank 32. Table \ref{tab:rankingPoints16-32-64RatioTo32} shows some statistics. The ranking point ratios at rank 16 and 64  to rank 32 are nearly 1.6 and 0.6, respectively.

\begin{table}[h]
\caption{Ranking points ratio to rank 32}
\label{tab:rankingPoints16-32-64RatioTo32}
\begin{center}
\begin{tabular}{cccc}\hline
& $16$ &$32$ &$64$ \\ \hline \hline
Expected&{\bf 1.9286}&{\bf 1}&{\bf 0.5159}\\ \hline
2017.03.20&1.9246&1&0.6135 \\ \hline
Maximum&{\bf 1.9737}   & 1  &  0.7056\\ \hline
Mean& 1.6445   & 1  & 0.6166\\ \hline
Minimum& 1.3699  &  1  &  {\bf 0.5190}\\ \hline
Std$*$& 0.1288   &      0  &  0.0330\\ \hline
\end{tabular}
\vspace{5pt}

$*$ Std: standard deviation
\end{center}
\end{table}

\subsection{Winning probability and ranking point ratio}

The previous section shows that the ratio between ranking points are essential value.

\begin{figure}[h]
\begin{center}
	\includegraphics[width=1.0\columnwidth ,clip]{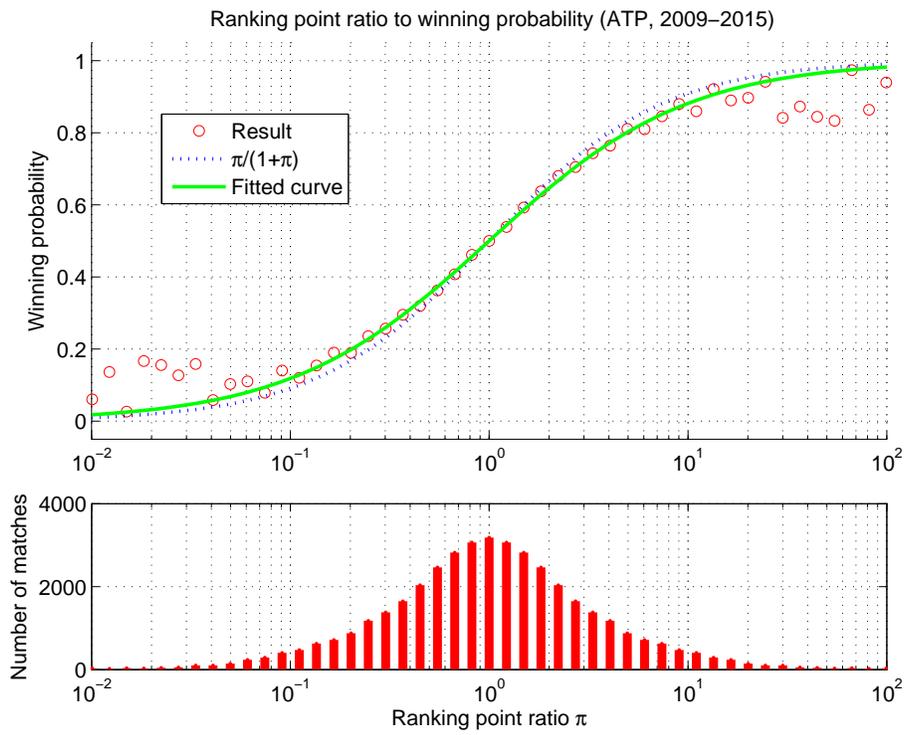}
	\caption{Ranking point ratio to winning probability (ATP)}
	\label{fig:atpResult}
\end{center}
\end{figure}

Figure \ref{fig:atpResult} shows the match result from 2009 to 2015.
It includes about 20000 matches in ATP world tour, Davis Cup, and Olympic Games.  
The horizontal axis is the ratio between players, $\pi_{i,j}=r_i/r_j, ~ (r_i, r_j$ are the ranking point of the players $i$ and $ j$.
The vertical axes of the op and the bottom figures are the results and the number of matches, respectively.
Match results are excluded if at least one player has 0 ranking points.

The ratio is from $0.1$ to $10$ in the major part of the matches.
A model
 $\hat{p}_{i,j}=\displaystyle \frac{\pi_{i,j}}{1+\pi_{i,j}}$ gives an approximation, however, the winning probability of higher-ranked players is overestimated.

Here, assume that a model
\begin{equation}
\label{eqn:defpij}
\hat{p}_{i,j}=\frac{\pi_{i,j}^\alpha}{1+\pi_{i,j}^ \alpha}, ~~\alpha>0,
\end{equation}
and find the parameter $\alpha$ that minimizes the difference between the match results $w$ and the predicted winning probability $\hat{p}$, i.e., minimize 
\begin{equation}
E^2=\frac{1}{number~of~matches}\sum_{for~all~ matches}(w_{i,j}-\hat{p}_{i,j})^2.
\end{equation}
$\alpha=0.8722$ gives the minimum $E^2=0.2052$. 
The bold line in the figure illustrates (\ref{eqn:defpij}).
As a comparison, the simplest prediction such as ``the player with higher ranking will win" gives $E^2=0.3227$.

\begin{figure}[h]
\begin{center}
	\includegraphics[width=1.0\columnwidth ,clip]{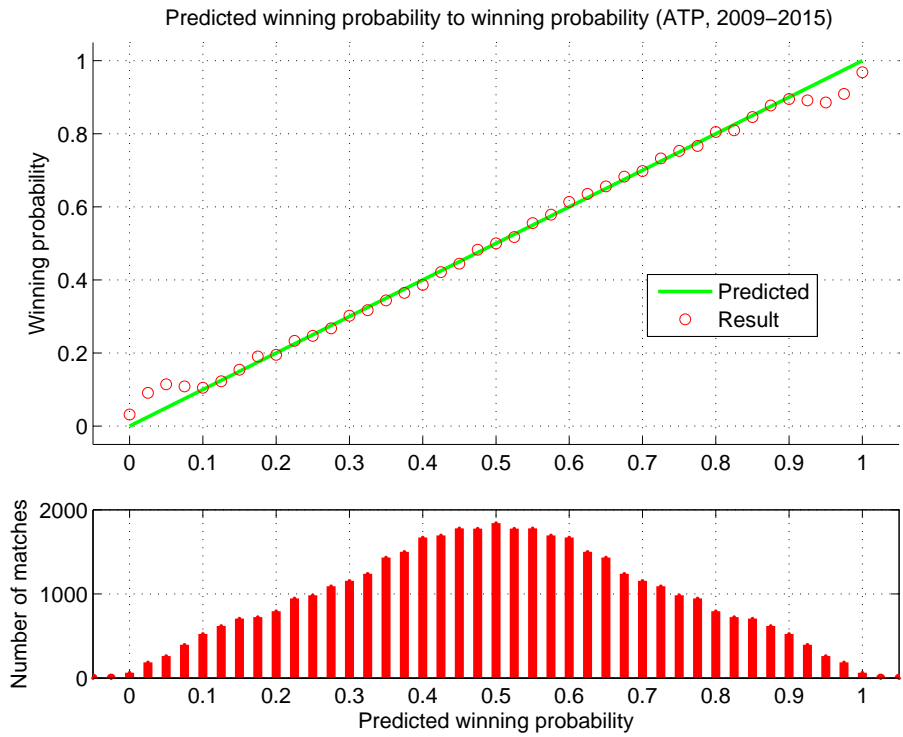}
	\caption{Predicted winning probability to result(ATP)}
	\label{fig:atpResultLinear}
\end{center}
\end{figure}

The horizontal axis is replaced with the predicted winning probability defined in (\ref{eqn:defpij}) in Figure \ref{fig:atpResultLinear}. 

The similar analysis are done for about 3400 match results in 2016 and 2017.
The result is $\alpha=0.8667$ and $E^2=0.2065$.
This shows that (\ref{eqn:defpij}) gives an consistent prediction model in recent years.

\section{Conclusion}
This paper shows that the following hypothesis can be draws from the definition of ATP ranking from 2009.
\begin{itemize}
\item ATP ranking point, the official ranking point system of the professional tennis players, can predict winning probability with the model  \begin{equation}
\hat{p}_{i,j}=\frac{\pi_{i,j}^\alpha}{1+\pi_{i,j}^\alpha},~~ \alpha=0.8722,
\end{equation}
where $\pi_{i,j}=r_{i}/r_{j}$ is the ranking point ratio between players $i$ and $j$.
\item This can be realized by consistent design of consistent overall tournament structure design, including the tournament categories, the winner's ranking points, seeding design, and so on. 
\end{itemize}

The hypothesis is validated using over 23000 match results from 2009 to 2017.
The match result shows that the prediction model can explain the match result in wide range of the ranking point ratio between players.

\end{document}